\documentclass[12pt,prl,aps,notitlepage]{revtex4}

\usepackage{slashed}
\usepackage{stmaryrd}
\usepackage{color}
\usepackage{mathrsfs}
\usepackage{yfonts}
\usepackage{rotating}
\usepackage{graphicx} 
\usepackage{amsfonts}
\usepackage{amsmath}
\usepackage{mathrsfs}
\usepackage[american]{babel}
\usepackage{wasysym}

\newcommand{\bea}{\begin{eqnarray}}
\newcommand{\eea}{\end{eqnarray}}
\newcommand{\beq}{\begin{equation}}
\newcommand{\eeq}{\end{equation}}

\everymath{\displaystyle}

\begin{document}
\title{Black Hole Quantum Vacuum Polarization in Higher Dimensions}
\author{Antonino Flachi}
\email{flachi@phys-h.keio.ac.jp}
\affiliation{Department of Physics, and Research and Education Center for Natural Sciences, Keio University, Hiyoshi 4-1-1, Yokohama, Kanagawa 223-8521, Japan
\,\,\,}
\author{Gon\c{c}alo M. Quinta}
\email{goncalo.quinta@ist.utl.pt}
\affiliation{Centro Multidisciplinar de Astrof\'{\i}sica, CENTRA,
Departamento de F\'{\i}sica, Instituto Superior T\'ecnico - IST,
Universidade de Lisboa - UL, Avenida Rovisco Pais 1,
1049-001 Lisboa, Portugal\,\,\,\,}
\author{Jos\'{e} P. S. Lemos}
\email{joselemos@ist.utl.pt}
\affiliation{Centro Multidisciplinar de Astrof\'{\i}sica, CENTRA,
Departamento de F\'{\i}sica, Instituto Superior T\'ecnico - IST,
Universidade de Lisboa - UL, Avenida Rovisco Pais 1,
1049-001 Lisboa, Portugal\,\,\,}

\begin{abstract}
The goal of this paper is to extend to higher dimensionality the methods and computations of vacuum polarization effects in black hole spacetimes. We focus our attention on the case of five dimensional Schwarzschild-Tangherlini black holes, for which we adapt the general method initially developed by Candelas and later refined by Anderson and others. We make use of point splitting regularization and of the WKB approximation to extract the divergences occuring in the coincidence limit of the Green function and, after calculating the counter-terms using the Schwinger - De Witt expansion, we explicitly prove the cancellation of the divergences and the regularity of the vacuum polarization once counter-terms are added up. We finally handle numerically the renormalized expression of the vacuum polarization. As a check on the method we also prove the regularity of the vacuum polarization in the six dimensional case in the large mass limit.
\end{abstract}
\maketitle

\newpage
\section{Introduction} 

The scalar vacuum polarization is the simplest way to probe quantum activity at the semiclassical level near a black hole. This quantity gives indication about symmetry breaking and it is usually regarded as preparatory to the more involved calculations of the quantum energy-momentum tensor, {\it i.e.} of the R.H.S. of the semiclassical Einstein equations. Chronologically, Candelas was the first to challenge the calculation of $\langle \phi^2 \rangle$ on the background of a four dimensional Schwarzschild black hole and in Ref.~\cite{Candelas:1980zt} he obtained an expression valid at the horizon. His method relied on point splitting regularization 
\cite{dewitt,Christensen:1976vb} and provided the main technical ingredients necessary for addressing similar calculations in other static black hole geometries. Two years later Page showed how to compute the same quantity outside the horizon by means of approximation \cite{Page:1982fm} and obtained a remarkably simple formula, $\langle \phi^2 \rangle = (T_{loc}^2 - T_{acc}^2)/12 + \mbox{\it corrections}$, with $T_{loc}=(1-2m/r)^{-1/2}/(8\pi m)$ being the locally red-shifted temperature and $T_{acc}=m/(1-2m/r)^{1/2}/(2\pi^2 r^2)$ the acceleration temperature corresponding to the local value of the acceleration ($m$ represents the black hole mass). The interpretation of this expression is that $\langle \phi^2 \rangle$ divides into a real particle contribution, proportional to $T_{loc}^2$, and a pure vacuum polarization contribution, proportional to $T_{acc}^2$. Curvature effects make up the corrections which were found, to some surprise, to be small \cite{Candelas:1984pg}. 
Subsequent work managed to obtain an expression valid outside the horizon and helped settle some of the details \cite{Fawcett:1981fw, Fawcett:1983dk,Candelas:1984pg}. Finally, Candelas and Jensen obtained $\langle \phi^2 \rangle$ inside the event horizon \cite{Candelas:1985ip}. An adaptation of these early calculations to general four dimensional, static, spherically symmetric and asymptotically flat black holes came some years later \cite{Anderson:1989vg}. 
Since the early calculations of Refs.~\cite{Fawcett:1981fw, Fawcett:1983dk,Candelas:1984pg}, various explicit cases have been considered and many works juxtaposed and followed those initial results, extending the calculations to various cases. A longer list of early references can be found in Ref.~\cite{frolovnovikov,parkertoms}. 

Two cases have received somewhat less attention and remain as outstanding problems. The first is the case of rotating black holes that proves to be technically very difficult, for which only partial results have been obtained (See, for example, Refs.~\cite{Frolov:1982pi,Duffy:2005mz,Belokogne:2014ysa,Cvetic:2015cca,Cvetic:2014eka}). The other case, relevant to the present work, is that of higher dimensional black holes that, also, has received only marginal attention. The first example of the sort is that of Ref.~\cite{Frolov:1989rv}, where an exact expression of $\langle \phi^2 \rangle$ at the event horizon of a five dimensional Schwarzschild black hole was derived. Ref.~\cite{Shiraishi:1993ti} calculated the vacuum polarization outside a five dimensional AdS black hole in a modified theory of gravity.
Refs.~\cite{Decanini:2005eg,Thompson:2008bk} have discussed issues related to renormalization. Ref.~\cite{Breen:2015hwa} appeared more recently and reported a calculation of $\langle \phi^2 \rangle$ for the four dimensional section of a brane-world black hole. 

In this paper we continue the analysis of quantum vacuum polarization around higher dimensional black holes and present the results for the vacuum polarization outside a $D=5$ Schwarzschild-Tangherlini black hole. One of the reasons to focus on the specific example of five dimensions is that the calculation becomes increasingly more cumbersome for higher dimensions and having a way to analyze the lower end of the dimensionality spectrum may provide a guide for addressing the problem in more general cases. As a check on our method we also prove the renormalization of the vacuum polarization for $D=6$ in the large mass limit where we can compare our results with those of Ref.~\cite{Thompson:2008bk} in the same limit.

Our method, described in the next section, adapts that of Refs.~\cite{Candelas:1984pg,Anderson:1989vg,Flachi:2008sr} to the present case and makes use of the WKB approximation along with some re-arrangements that are useful for the subsequent numerical evaluation.  
In fact, while the method of Refs.~\cite{Candelas:1984pg,Anderson:1989vg,Flachi:2008sr} formally extends to higher dimensions in a  straightforward manner, difficulties appear at various stages of the calculations. First of all, in higher dimensions it is necessary to push the WKB expansion to increasingly high orders for reasons related to both convergence and renormalization. Secondly, only very few explicit results are available beyond four dimensions for the counter-terms. In fact, the only result we are aware of is that of Ref.~\cite{Thompson:2008bk} where an explicit formula for $D=6$ is given. We are not aware of any explicit result in $D=5$ or $D>6$. {A simpler approach would be to} skip the verification of the regularity, start from the assumption that the result for the vacuum polarization is renormalizable and simply drop the divergences from the result. However, considering the algebraic complexity of the calculation, explicitly proving the cancellation of the divergences offers a very non-trivial check on the results. Practically, this comes at the price of having to compute the counter-terms in higher dimensions, which is also non-trivial. Finally, in higher dimensions it is indispensable to carry out the algebraically more complex parts of the calculations using computer symbolic algebra. Automatisation of the calculation too becomes increasingly more messy as the dimensionality increases and explicit examples should be regarded as a useful guide towards addressing the full higher dimensional problem.

\section{The Method}

Our calculation deals with the exterior region of a 
 $D$-dimensional Schwarzschild-Tangherlini black hole that, after euclideanization, is described by the following metric
\bea
ds^2 = f( r ) d\tau^2 + f^{-1}( r ) {dr^2} + r^{ N + 1} d\Omega_{N+1}^2\,,
\label{LineEle}
\eea
where $\tau$ is the Euclidean time direction with period $\beta$, $r$ is the radial coordinate, $\Omega_{N+1}$ is the D-dimensional solid angle,  $N=D-3$, and $f(r) = 1 - (2 M_{BH}/r)^{N}$. 
We remind that we focus on $D=5$ and $D=6$ in the next sections.

The vacuum polarization in the Hartle-Hawking state can be expressed as the renormalized coincidence limit of the Euclidean Green function $G_E(x,x')$, which, by definition, satisfies the differential equation
\beq
\left(\Box - m^2 -\xi R \right) G_E(x,x') = - {\delta^{(D)} (x-x') \over \sqrt{g}}\,,
\eeq
where $\delta^{(D)} (x-x')$ is the $D$-dimensional Dirac delta function, $R$ is the Ricci curvature scalar, $g$ is the determinant of the metric and $m$ is the mass of the associated scalar field. Prior to renormalization, it can be written as
\bea
G_E(x,x') &=& {1\over \beta} 
\sum_{n=-\infty}^{\infty} e^{i \omega_n (\tau - \tau')}
\sum_{l=0}^{\infty} 
k_l\; C^{(N/2)}_l(\Omega\cdot\Omega')\times
G_{ln}\left(r,r'\right)
\label{GreenExp}
\eea
where $k_l = {\Gamma(1+{N/ 2}) }(1 +{2l/ N})/(2\pi^{1+N/2})$ and $\omega_n = 2\pi n/\beta$, with $\beta$ being the inverse of the black hole temperature $T$. The Gegenbauer polynomials $C^{(N/2)}_l(\Omega\cdot\Omega')$ generalize to higher dimensions the Legendre functions and result from the summation over the azimuthal quantum numbers of the hyper-spherical harmonics in $D$ dimensions (relevant formulae can be found in Ref.~\cite{grad}). The radial part of the Green function can be written in terms of the two independent solutions $\chi_{nl}^\pm(r)$ of the homogeneous radial wave equation
\bea
{d^2 \chi_{nl}^\pm(r) \over dr^2}
+
\left({N+1\over r} + {f'\over f}\right)
{d \chi_{nl}^\pm( r) \over dr}
-
\left({\omega_n^2\over f^2} + {l(l+N) \over f r^2} +{m^2+\xi R \over f}\right){\chi_{nl}^\pm¨}(r) &=& 0\,,
\label{HomogEq}
\eea
giving
\bea
G_{ln}\left(r,r'\right) = 
{1\over r^{N+1} f}\;
{\chi_{nl}^+(r_{<})
\chi_{nl}^-(r_{>})
\over 
\langle
\chi_{nl}^+,\chi_{nl}^-
\rangle
}
\label{GreenF}
\eea
where $\langle\chi_{nl}^+,\chi_{nl}^-\rangle$ indicates the wronskian of the two solutions. 

It is well known that direct numerical evaluation of (\ref{GreenExp}) is impeded by the diverging nature of the coincidence limit. To bypass the problem, it is customary to use point-splitting and take the coincidence limit along all directions but one. Here, we choose to use point-splitting along the time direction. We then use the WKB approximation to explicitly extract the divergences. Subsequently, we compute the counter-terms using the Schwinger - De Witt expansion and subtract these from the regulated expression of the unrenormalized vacuum polarization. This way of proceeding is nothing but a generalization of the method developed by Candelas and later refined by Anderson.

A convenient WKB ansatz for the solution is
\bea
\chi_{nl}^\pm(r) =  {\exp\left({\pm \int_{r_s}^{r} \Psi(z) {dz \over f(z)}}\right) \over  \sqrt{r^{1+N} \Psi}}\,,
\label{AnsatzWKB}
\eea
where we have added the term ${1/ f}$ in the exponential in order to make explicit the asymptotic behavior of the solutions, as it can be checked. Obtaining $\Psi$ to first order in the WKB approximation is sufficient to extract the divergences in the four-dimensional case treated by Anderson, while higher dimensional cases it is necessary to increase the order of approximation. {By inserting Eq.~(\ref{AnsatzWKB}) in Eq.~(\ref{HomogEq}), one may expand $\Psi$ in general as} 
\bea
\Psi^{-1} = \Phi^{-1/2}\left(1 + {\delta_{1}\Phi} + {\delta_{2}\Phi}  + \cdots \right)
\label{PsiApprox}
\eea
with the dots indicating higher order WKB approximants. {The following expressions can be obtained for the first three terms}
\bea
\Phi &=& {f\over r^2} \left[\left(l+ {N\over 2}\right)^{2}- {N^2 \over 4}\right] + \Omega_n^2\,, ~~~~~
\label{Phi}
\\
\delta_1\Phi &=& \left(
{5 f^2\over 32}{\Phi^{'2}\over \Phi^3}-{f^2\over 8}{\Phi''\over \Phi^2} - {f f' \over 8}{\Phi'\over \Phi^2}
\right)\,,
\label{delta1}
\\
\delta_2\Phi &=& 
\frac{11 {a_1}^2 \Phi^{'2}}{32 \Phi ^4}-\frac{{a_1}^2 \Phi ''}{8
   \Phi ^3}+\frac{17 {a_1} {a_2} \Phi^{'3}}{32 \Phi
   ^5}-\frac{{a_1} {a_2} \Phi ' \Phi ''}{4 \Phi ^4}-\frac{{a_1} {a_3}
   \Phi ^{'''}}{4 \Phi ^3}-\frac{25 {a_1} {a_3} \Phi^{'3}}{32 \Phi
   ^5}
   \nonumber\\
&+&   
   \frac{19 {a_1} {a_3} \Phi ' \Phi ''}{16 \Phi ^4}+\frac{27 {a_2}^2
  \Phi^{'4}}{128 \Phi ^6}-\frac{{a_2}^2\Phi^{'2} \Phi
   ''}{8 \Phi ^5}-\frac{51 {a_2} {a_3} \Phi^{'4}}{64 \Phi
   ^6}-\frac{{a_2} {a_3} \Phi^{''2}}{8 \Phi ^4}
   \nonumber\\
&+&   
   \frac{41 {a_2}
   {a_3} \Phi^{'2} \Phi ''}{32 \Phi ^5}-\frac{{a_2} {a_3} \Phi
   ^{'''} \Phi '}{4 \Phi ^4}+\frac{75 {a_3}^2 \Phi^{'4}}{128 \Phi
   ^6}-\frac{{a_3}^2 \Phi^{'''}}{8 \Phi ^3}+\frac{15 {a_3}^2 
   \Phi^{''2}}{32 \Phi ^4}
   \nonumber\\
&-&   
   \frac{45 {a_3}^2\Phi^{'2} \Phi ''}{32 \Phi
   ^5}+\frac{{a_3}^2 \Phi ^{'''} \Phi '}{2 \Phi ^4}\,,
   \label{delta2}
\eea
where 
\bea
\Omega_n^2 &=& \omega_n^2 + \mu^2 f, \\
\mu^2 &=& m^2 + \xi R +{N^2-1\over 4r^2} f + {N+1\over 2 r}f'.
\eea
Without loss of generality, in the semiclassical regime for the Schwarzschild-Tangherlini geometry, we may drop the minimally coupled part and set $\xi=0$. However, we retain such a term for two reasons. 
First of all, our analytical computation is valid for more general, other than black holes, spherically symmetric spacetimes. In this case such non-minimally coupled terms are present and generate additional diverging contributions to the vacuum polarization that should also be compensated by counter-terms. Secondly, once these terms are added additional cancellations should take place with the counter-terms providing another stringent check on the results. 

In the following we will re-group all the WKB approximants and define, for notational convenience, $\delta \Phi = \delta_1\Phi + \delta_2\Phi + \ldots$. Using Eq.~(\ref{GreenExp}) and Eq.~(\ref{GreenF}) we can express the coincidence limit of the Green function as
\bea
G_E(x,x) &=& 
{1\over \sqrt{4\pi} \beta} {1\over (\pi r^2)^{1+N\over 2}}
{\Gamma\left(1+{N\over 2}\right) \over \Gamma\left(1+{N}\right)}\times 
\sum_{n,l} 
\left(l+1\right)^{\{N-1\}} 
\left[\left(l +{N\over 2}\right) 
\Psi^{-1}
\right],
\label{GreenCoinc}
\eea
where the quantity $\left(b \right)^{\{a\}}$ defines the Pochhammer symbol
\beq
\left(b \right)^{\{a\}} = {\Gamma(b+a) \over \Gamma(b)}  \,.
\eeq
It is usual to subtract off the divergences in $l$ by adding terms proportional to $\delta(\tau - \tau')$ that do not alter the Green function and remove the nonphysical divergences. We will carry out this procedure at the end by dropping the $n$-independent contributions diverging for large $l$, corresponding to the above mentioned subtraction.  
We should note that Eq.~(\ref{GreenCoinc}) is formal as we have not yet added the counter-terms. 

It is convenient to rearrange the {unrenormalized} Green function as $G_E(x,x) = G_{0} + \delta G$ 
separating the terms containing the divergences,
\bea
G_{0}&=&
{1\over \sqrt{4\pi} \beta} {1\over (\pi r^2)^{1+N\over 2}}
{\Gamma\left(1+{N\over 2}\right) \over \Gamma\left(1+{N}\right)}
\times
2\sum_{n=1}^{\infty}{\sum_{l=0}^{\infty}}^{\prime}
J^{(D)}_{n,l}
\eea
from those that are regular by construction, 
\bea
\delta G &=&
{1\over \sqrt{4\pi} \beta} {1\over (\pi r^2)^{1+N\over 2}}
{\Gamma\left(1+{N\over 2}\right) \over \Gamma\left(1+{N}\right)}
\times
\left\{
{\sum_{l=0}^{\infty}}^{\prime} 
J^{(D)}_{0,l}
+\sum_{n=-\infty}^{\infty}
\sum_{l=0}^\infty 
\left(
\tilde J^{(D)}_{n,l}
-
J^{(D)}_{n,l}
\right)\right\}\,.
\eea
In the formulae above, we have defined $J^{(D)}_{n,l} = \left(l+1\right)^{\{N-1\}} \left(l+{N/ 2} \right){(1 + {\delta\Phi})/ \sqrt{\Phi}}$, that is the WKB approximation for $\tilde J^{(D)}_{n,l} = \left(l+1\right)^{\{N-1\}} \left(l+{N/ 2} \right){/ {\Psi}}$. 
The primes in the summations are a reminder that the diverging contributions at large $l$ have to be dropped. The second term in $\delta G$ is nothing but the reminder of the WKB approximation and it can be directly evaluated numerically. Including higher order WKB approximants makes this difference smaller.

A general procedure to extract the divergences follows by making use of the Abel-Plana summation formula that allows turning the summation over $l$ into an integral expression. Defining $j_n^{(D)}(l) \equiv J^{(D)}_{n,l}$, {in order to highlight the $l$ dependence}, we find
\bea
G_{0} =
{1/( \sqrt{\pi} \beta)\over (\pi r^2)^{1+N\over 2}}
{\Gamma\left(1+{N\over 2}\right) \over \Gamma\left(1+{N}\right)}
\sum_{n=1}^{\infty}
\left[
{j_n^{(D)}(0)\over 2}
+
 \int_0^\infty j_n^{(D)}(x) dx
+
i \int_0^\infty {j_n^{(D)}(i x) -j_n^{(D)}(-i x) \over e^{2\pi x}-1 } dx
\right].~~~
\label{ap}
\eea
The first term above can can be computed using zeta-regularization techniques, while the second integral can be performed analytically with the results expressed in terms of hypergeometric functions. The third integral can be computed by approximation. 
The divergences can be extracted, after the sums and integrals are calculated, by expanding for large frequencies. Already for $D=5$ results becomes very lengthy and we will not report here any formula, although below we will explicitly give the divergences that are generated from each of the above terms. 

To arrive at the renormalized vacuum polarization we need to add appropriate counter-terms. These have been computed by Christensen in Ref.~(\cite{Christensen:1976vb}) in four-dimensional spherically symmetric spacetimes using the short-distance (Schwinger - De Witt) approximation of the propagator in curved space. His results have made the subtraction procedure straightforward in most cases. In higher dimensions, Ref.~\cite{Thompson:2008bk} has reported explicit results for the counter-terms in six-dimensional spherically symmetric spacetimes. We are not aware of any relevant results in $D=5$ or for $D>6$. Here we compute the counter-terms independently and aside for reproducing Christensen's four-dimensional results as a check, we obtain new expressions in five dimensions. We also calculate them in the six dimensional large mass limit, reproducing the counter-terms calculated in Ref.~\cite{Thompson:2008bk}. 

\section{Five dimensions}

As discussed in the previous section, the renormalized vacuum polarization consists of two parts, $G_0$ and $\delta G$, plus the counter-terms. The contribution $\delta G$ is regular by construction and it contains the reminder of the WKB approximation. It can be calculated numerically once an explicit form for the metric is chosen. The contribution $G_0$ can be evaluated analytically in the way we have described in the previous section with the divergences extracted from a high frequency expansion. For the five dimensional case, all the divergent contributions are encoded in the first order WKB approximant $\delta_1\Phi$. However, here we will retain terms up to second order, i.e. $\delta \Phi = \delta_1 \Phi + \delta_2 \Phi$, as this will improve the numerical approximation. The result we find for $G_0$, with the explicit behaviour at large $l$ subtracted, is
\begin{widetext}
\bea
G_0 &=&
{1 \over 2 \pi^2 r^3 \beta} \sum^{\infty}_{n=1} \sum^{\infty}_{l=0} \left\{ {(l+1)^2 \Psi^{-1}} - {(l+1)r \over \sqrt{f}} \right\} \nonumber \\
&=&
{1 \over 2 \pi^2 r^3 \beta} 
\sum_{n=1}^{\infty}
\left\{
\textfrak{Z}_{reg}^{(1)}+\sum_{k=5}^{13} \alpha_k \textfrak{Z}^{(k)}
+\textfrak{J}_{reg}+\textfrak{P}_{reg}
\right\} + G_{div.},
\label{eqsdf}
\eea
where
\bea
\textfrak{Z}^{(k)}
=
\sum_{n=1}^\infty \left(
\omega_n^2 + f \mu^2
\right)^{-k/2}, ~~~~~
\textfrak{Z}^{(1)}_{reg}
=
{1\over 2} \sum_{n=1}^\infty \left[
\left(
\omega_n^2 + f \mu^2
\right)^{-1/2}
-
{1\over \omega_n}
\right],
\eea
and
\bea
\alpha_5&=&-\frac{3 f^2 \left(-f^{'2} x^2+f^{'} f r+3 f^2\right)}{32 r^4},\nonumber\\
\alpha_7&=& 
\frac{3 f^3 \left(-4 f^{'3} r^3+33 f^{'2} f r^2+30 f' f^2 r-105 f^3\right)}{256 r^6},\nonumber \\
\alpha_{9} &=& \frac{9 f^4 \left(11 f^{'4} r^4-78 f^{'3} f r^3+13 f^{'2} f^2 r^2+606 f^{'} f^3 r+525 f^4 \right)}{1024 r^8},\nonumber\\
\alpha_{11} &=& 
-\frac{27 f^6 (f' r+f)^2 \left(-101 f^{'2} r^2+361 f' f r+693 f^2\right)}{4096 r^{10}},\nonumber\\
\alpha_{13} &=& \frac{93555 f^8 (f' r+f)^4}{65536 r^{12}}.\nonumber
\eea
The expression for $\textfrak{J}_{reg}$ can be calculated explicitly in terms of hypergeometric functions. However, the result is very long and will not be reported here, although it can be obtained more or less straightforwardly using a symbolic manipulation program. 
It should also be noted that the upper bound on the sum over $k$ in (\ref{eqsdf}) reflects the order of the WKB approximation, {since higher orders} increases the numbers of zeta functions to be added up. {The fact that the WKB approximation part of the vacuum polarization can be expressed as a series} of such zeta functions was already noted in Ref.~\cite{Flachi:2008sr}. {Rather than showing the explicit results for the integrals,} it is more instructive to explicitly report the diverging contributions generated by each term, which we subtract in order to regulate each expression:
\bea
\textfrak{J}_{reg} =
\int_0^\infty  j_n^{(5)}(x) dx
&-&
\left\{
-{1\over 3 \omega_n}
+{r^3 \omega_n^2\over 2 f^{3/2}}
\left(
1+\ln\left(
{r \omega_n \over 2 \sqrt{f}}
\right)
\right)
+
\tilde a_1
\left(
1+\ln\left(
{r \omega_n \over 2 \sqrt{f}}
\right)\right)
\right\}~~
\eea
where
$$
\tilde a_1 = {r^3\over 8 \sqrt{f}} \left( m^2 - [a_1] +{f'\over r} - {f^{'2} \over 4 f} + {f^{''}\over 3} \right),
$$
with $[a_1] = - (\xi - 1/6) R$ being the first non-trivial heat-kernel coefficient (for $m=0$) and
\bea
\textfrak{P}_{reg} =
i \int_0^\infty {j^{(5)}_n(i x) -j^{(5)}_n (-i x) \over e^{2\pi x}-1 } dx
+{1\over 6\omega_n}.
\eea
The terms that we have subtracted to define the regulated quantities correspond to the divergent contributions that each term generates. {Adding all these contributions we are left with the divergent part of the Green function}
\beq
G_{div} = {1 \over 16 \pi^2 f^{3/2} \beta}\sum^{\infty}_{n=1}
\left[{4 \omega_n^2\over f }
+
\left(m^2 - [a_1] +{f'\over r} - {f^{'2} \over 4 f} + {f^{''}\over 3} \right)
\right]\left(
1+\ln\left(
{r \omega_n \over 2 \sqrt{f}}
\right)\right)\,,
\eeq
which should be compensated by the counter-terms, leaving a perfectly regular expression for the renormalized vacuum polarization. It is worth noticing that the divergences proportional to the inverse frequency $1/\omega_n$ cancel against each other, i.e. in five dimensions the counter-terms do not generate such divergences.

The computation of the counter-terms is also lengthly, but with the help of computer symbolic manipulation it can be carried out directly. The general expression for the adiabatic Schwinger-DeWitt expansion of the propagator in general dimensions is already known from Christensen's work \cite{Christensen:1976vb} (see also Refs.~\cite{Frolov:1989rv,Thompson:2008bk}) that we report for the convenience of the reader:
\bea
G_{SDW} (x,x') 
= {i \pi \Delta_{vvm}(x,x') \over (4 \pi i)^{D/2}}
\sum_{k\geq 0}
a_k(x,x')
\left( -{\partial \over \partial m^2} \right)^k
\left( -{z \over  2 i m^2} \right)^{1-D/2} H_{D/2-1}^{(2)}(z),
\eea
where $\Delta_{vvm}(x,x')$ is the Van Vleck-Morette determinant, $z^2 = - 2 m^2 \sigma(x,x')$ with $2\sigma(x,x') = s^2(x,x')$  being the square of the geodesic distance, and $H_{D/2-1}^{(2)}(z)$ is a Hankel function of second kind. From the above expression, the procedure to extract the counter-terms consists in choosing a separation between the points and expanding. Splitting along the time direction, for which
\bea
\sigma = {f\over 2} \varepsilon^2 -{f f^{'2}\over 96} \varepsilon^4 + O(\varepsilon^6),
\eea
computing the various quantities involved, and expanding for small separation, gives the result
\bea
\langle \phi^2 \rangle_{div}
\equiv G_{SDW} (x,x)  = \lim_{\varepsilon \to 0} \left\{
-{1\over 8 \pi^2 f^{3/2} \varepsilon^3} 
-{1\over 16 \pi^2 \sqrt{f}\varepsilon }
\left(
m^2 - [a_1] + 
{f'\over 4r}
-{f^{'2}\over 16}
+
{f'' \over 12}
\right)\right\}~
\eea
for the divergent part of the vacuum polarization that needs to be subtracted. In order to perform the subtraction of the above counter-terms, we first need to re-express the quantity above in terms of summation over $\omega_n$. This can be done by adapting the general procedure outlined in \cite{Howard:1984,Anderson:1989vg}, using the Abel-Plana rearrangement for the function $\left(a^2 \omega_n^2 \right)^s \cos\left(\omega_n \varepsilon\right)$, taking the first derivative with respect to $s$ and then setting $s=0$. One then arrives at the following expression
\bea
\sum_{n=1}^\infty \log \left( a^2 \omega_n^2\right) \cos\left(\omega_n \varepsilon\right)
&=&
-{\pi \beta \over 2\pi \varepsilon}  + O(\varepsilon)\,.
\eea
Differentiating three times gives the relevant rearrangements for $1/\varepsilon^3$. It is then a matter of algebra to show that the counter-terms, $\langle \phi^2 \rangle_{div}$, precisely compensate the divergences $G_{div}$ that we have extracted using the WKB approximation.

The next part of the calculation involves the numerical evaluation of the renormalized vacuum polarization, given by
\beq
\langle \phi^2 \rangle_{ren} = (G_0 + \delta G) - \langle \phi^2 \rangle_{div}\,.
\eeq
The numerical procedure can be carried out in a straightforward manner. The only term that becomes computationally demanding is the remainder. A way to proceed in this case is to increase the order of the WKB approximation which renders this term small. In our numerical evaluation we have used up to third order approximation and ignored the reminder. Results are illustrated in Fig.~\ref{fig1} for various values of the parameters. As it should, results are regular at the horizon and we find no problem of convergence in the numerical approximation.

\begin{figure}[h]
\begin{center}
\unitlength=1mm
\unitlength=1mm
\begin{picture}(180,60)
   \includegraphics[height=5cm]{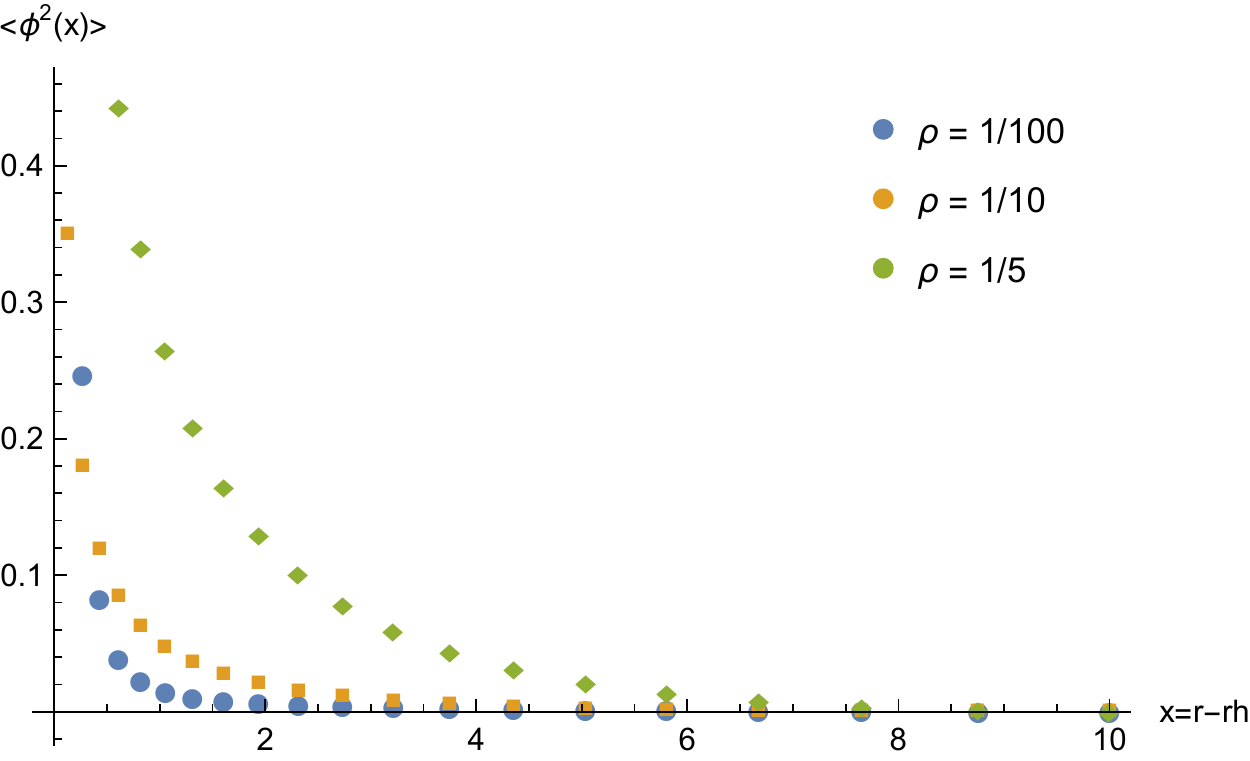}
   \includegraphics[height=5cm]{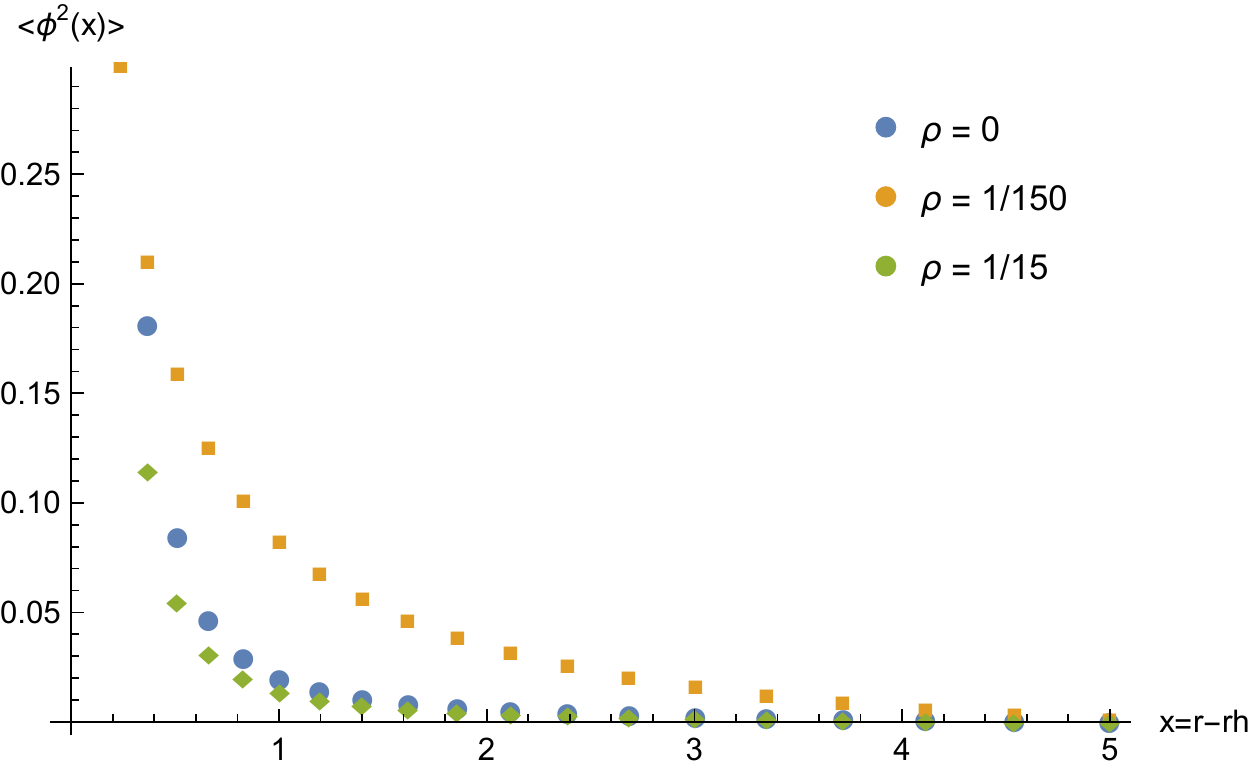}
\end{picture}
\end{center}
\caption{Profile of the renormalized vacuum polarization for $M_{BH}=5$ (left-panel) and $M_{BH}=15/2$ (right-panel), for various values of the parameter $\rho={m\over 2M_{BH}}$.}  
\label{fig1}
\end{figure}

\section{A check in six dimensions}

The six-dimensional case is computationally more demanding and extracting the divergences is not trivial. The reason for the increased complexity comes from the fact that higher order WKB terms give rise to additional ultraviolet divergent contributions and this leads to algebraically very cumbersome combinations of hypergeometric functions. Computing the counter-terms also becomes more difficult as the dimensionality increases. Some simplification can be achieved in the specific limit of large mass and here, as a check on the method, we have limited our analysis to this case. The procedure to extract the divergences is basically the same as in five dimensions and consists in operating on the Abel-Plana rearrangement of $G_0$. Keeping the first two leading terms in a large mass expansions we find the following diverging behaviour for the Green function:
\bea
G^{(6D)}_{div} \approx {1\over 64 \pi^3\beta} \sum^{\infty}_{n=1}{1 \over \omega_n }\left\{m^4 + m^2 \left[
\frac{4 \omega_n^2}{f}
+2\left(\frac{1}{6} - \xi \right)
\left(f''+{8f'\over r} +{12f\over r^2} -{12\over r^2} \right)
\right]\right\},
\eea
where $\approx$ reminds that we are considering only the first two terms in a large mass expansion. In extracting the divergences above all the terms above are generated by the second integral in the Abel-Plana rearrangement (\ref{ap}) in six dimensions. The form of the divergences can be quickly understood from the general form of the heat-kernel coefficients and dimensional analysis. It can be noted that the last term multiplying $m^2$ is the scalar curvature in six dimensions, as one might have guessed from dimensional analysis and from the general form of the heat-kernel coefficients. The counter-terms can be extracted similarly to what we did in five dimensions, giving, in the large mass limit,
\beq
\langle \phi^2 \rangle^{(6)}_{div} = \left({m^4 \over 64 \pi^3} - {[a_1] m^2 \over 32\pi^3}\right)\log\varepsilon  -  {m^2 \over 16 \pi^3 f}
\frac{1}{\varepsilon^2}\,.
\eeq
Using the expressions from Refs.~(\cite{Howard:1984,Anderson:1989vg}), relating $\log \varepsilon$ and $1/\varepsilon^2$ with sums of powers of $\omega_n$, it is easy to check that our result exactly compensates the divergences and agrees with the result of Ref.~\cite{Thompson:2008bk} in the same limit.

\section{Conclusions}

In this paper we have studied the renormalized vacuum polarization for a higher dimensional Schwarzschild-Tangherlini black hole. We have presented a general approach for computing the vacuum polarization and fully analyzed the problem in five dimensions. We have extracted the divergences using the WKB approximation and explicitly calculated the counter-terms proving the regularity of the result. Finally, we have evaluated the renormalized expression numerically. In six dimensions, we have limited ourselves to proving the regularity of the vacuum polarization in the large mass limit. We have too, in this case, extracted the divergences using the WKB approximation, calculated the counter-terms and explicitly verified the regularity. Our results for the counter-terms in the large mass limit coincide with those of Ref.~\cite{Thompson:2008bk}. Straightforward generalizations of the calculations of this paper include other black hole geometries with charge and cosmological constant in five dimensions. It should also be possible to relax the high mass approximation and work out in full detail along similar lines the six dimensional case with some effort. The most interesting and non-trivial generalization consists in increasing the number of dimensions for which the computations presented in this paper are a very useful warm-up. 

\section{Acknowledgements} We are grateful to R. Thompson for sharing with us his results for the six-dimensional case. This work is supported in part by the MEXT-Supported Program for the Strategic Research Foundation at Private Universities ``Topological Science'' (Grant No. S1511006) and by the Funda\c c\~ao para a Ci\^encia e Tecnologia of Portugal (FCT) through Projects No.~PEst-OE/FIS/UI0099/2015 and No.~SFRH/BD/92583/2013.

\end{widetext}

\newpage
\end{document}